\documentclass[epj]{svjour}

\usepackage{amsmath}
\usepackage{amssymb}
\usepackage[english]{babel}

\usepackage{ifpdf}
\ifpdf
\usepackage[pdftex]{graphicx}
\else
\usepackage[dvips]{graphicx}
\fi

\usepackage{wrapfig}

\usepackage{url}

\begin{document}


\title{A remarkable emergent property of spontaneous (amino acid content) symmetry breaking}

\author{R.A. Broglia\inst{1,2,3,4}}
\institute{Department of Physics, University of Milan, via Celoria 16, I--20133 Milan, Italy \and INFN, Milan Section, Italy \and The Niels Bohr Institute, University of Copenhagen, Blegdamsveg 17, DK--2100 Copenhagen, Denmark \and Foldless S.r.l. Via Valosa di Sopra, 9, I--20050, Monza(MB), Italy}


\abstract{Learning how proteins fold will hardly have any impact in the way conventional --- active site centered --- drugs are designed. On the other hand, this knowledge is proving instrumental in defining a new paradigm for the identification of drugs against any target protein: folding inhibition. Targeting folding renders drugs less prone to elicit spontaneous genetic mutations which in many cases, notably in connection with viruses like the Human Immunodeficiency Virus (HIV), can block therapeutic action. From the progress which has taken place during the last years in the understanding of the becoming of a protein, and how to read from the corresponding sequences the associated three--dimensional, biologically active, native structure, the idea of non--conventional (folding) inhibitors and thus of leads to eventual drugs to fight disease, arguably, without creating resistance, emerges as a distinct possibility.}

\maketitle

To the question\cite{Sch:44} ``what is life?'' one is forced to answer that life is not one but two things\cite{Dys:99}. Which ones ? Replication and metabolism. The molecules of DNA and RNA are responsible for the first function\cite{Wat:53,Wat:80}, proteins for the second\cite{San:52}. Because software (replication) is necessary a parasite of hardware (proteins), the becoming of a protein carries, to a large extent, the secret of life\cite{Mon:72}.

\begin{figure*}[htb!]
\begin{center}
\includegraphics[width=0.95\textwidth]{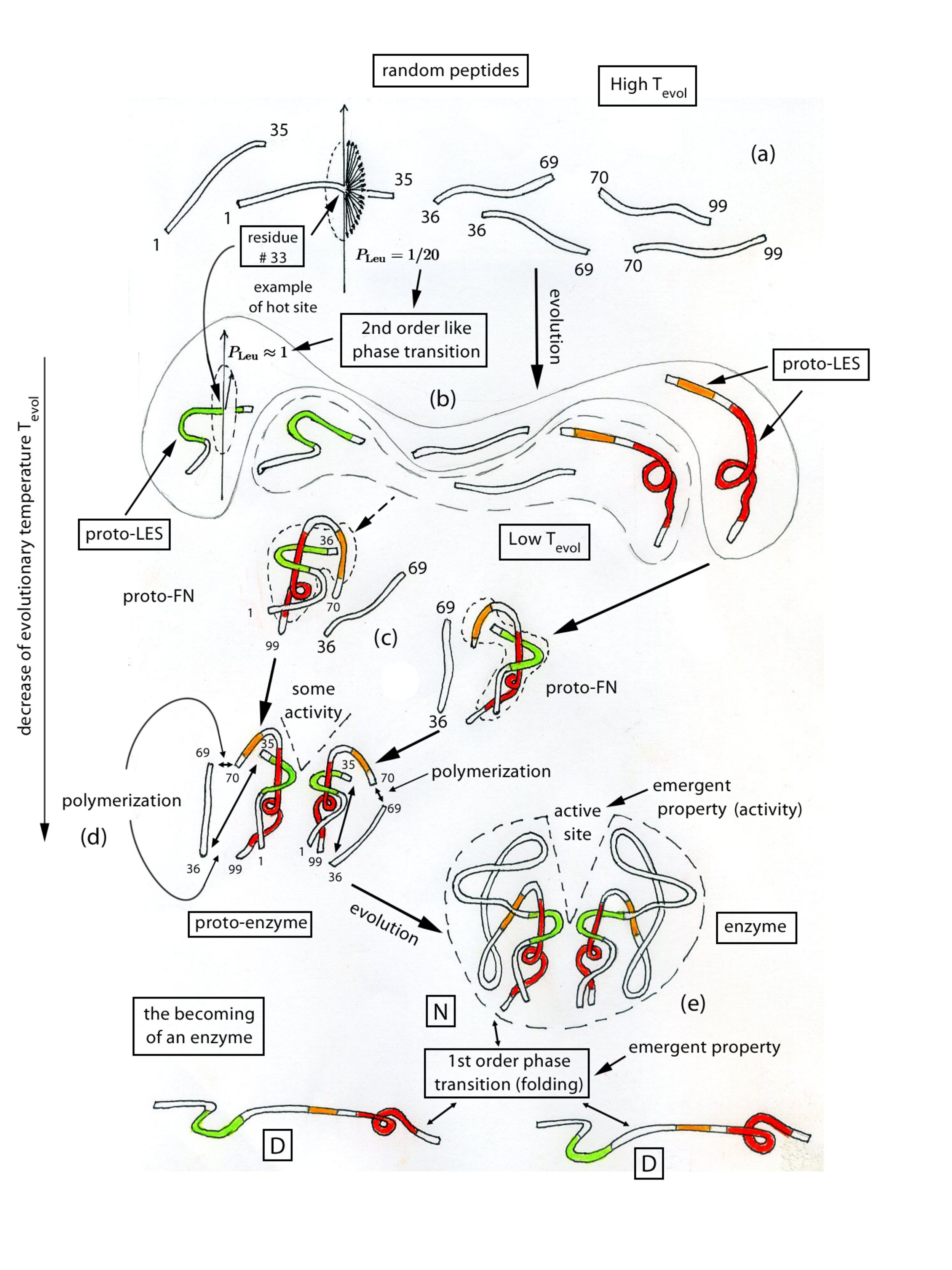}
\caption{}
\label{fig:protevo}
\end{center}
\end{figure*}

A possible scenario for this becoming suggests that, starting from random polypeptide chains (i.e. chains where the probability that a site is occupied by a given amino acid is $1/20$) containing some tens of amino acids (Fig. \ref{fig:protevo}(a)),
evolution rang a large number of all possible sequences until it clicked on a class of them containing few (4--6) strongly hydrophobic (``hot'') amino acids\footnote{In keeping with the fact that there are 20 different types of amino acids, one can associate with each site of the protein a quasispin (see e.g. \cite{And:58}) of value $19/2$, interpreting each projection as a given amino acid realization. In a random polymer, each projection is equally probable for any site, and no alignment is observed. Lowering the evolutionary temperature \cite{Sha:94,Mad:94} one finds that there exists a critical temperature (equivalent to the Curie temperature in the case of a ferromagnet and which can be simply calculated making use of the random energy model \cite{Der:81} ($E_C \rightarrow T_C$), see also \cite{Tia:98} and refs. therein) below which one observes quasispin alignment at specific sites (hot sites), an example of which is provided by the residue occupying site 33 (Leucine (Leu)) of the HIV--1--PR, a strongly hydrophobic, highly conserved amino acid which plays the role of hub \cite{Bar:02,Ven:02} (see also \cite{Bro:06} Fig. 39 (b) and Table VI) in the native conformation of this enzyme. Assuming that the high quasispin projections correspond to strongly hydrophobic amino acids, one essentially finds an alignment pointing along the (positive) quantization axis. That is, there is a priviledged orientation in quasispin space, and thus an associated spontaneous breaking of amino acid content symmetry, symmetry associated with the invariance of the Hamiltonian describing the interaction among amino acids with respect to amino acid occupancy of the different sites of the protein. This is similar to the spontaneous breaking of rotational symmetry associated with the ferromagnetic state below the Curie temperature, symmmetry respected by the original spin--spin Hamiltonian (see e.g. \cite{And:84,Ram:94} and \cite{Dub:10}.}, which induced the formation and provided the varied stability to Local Elementary Structure (LES)\cite{Tia:98,Bro:98,Bro:01b} (Fig. \ref{fig:protevo}(b)) which flicker in and out the native conformation\cite{Anf:73} (see also App. A). The segment of the polypeptide chain associated with a LES contains approximately 10--15 amino acids. This in keeping with the fact that segments of this length are able to fold in milliseconds, consistent with the fact that proteins containing about one hundred amino acids fold in times of the order of tens of milliseconds\cite{Fer:99}.

Strongly hydrophobic, highly conserved (hot) amino acids\cite{Tia:01,Bro:01a,Sha:96} inducing local structuring of the protein (see App. B) are responsible to a large extent for the selective interaction (molecular recognition) between a small group (2--4) of complementary\cite{Pau:40} (in the sense of left and right hands) LES (one or two per chain). When this group of LES docks, they give rise to the (postcritical) folding nucleus\cite{Abk:94} $(\textrm{FN})_\textrm{pc}$ (Fig. \ref{fig:protevo}(c)), which inevitably grows into a folded oligomer, that is, a unique three--dimensional structure with incipient specific biological functions, e.g. some amount of enzymatic activity\cite{Anf:73}. Polymerization (Fig. \ref{fig:protevo}(d)) gives rise to typical globular proteins (or folding domains), containing 100--120 amino acids\cite{Xu:97} (Fig. \ref{fig:protevo}(e)) (see App. C).

The setting in place, by evolution, of hot amino acids\cite{Tia:98}, that is the occupation of certain sites of the polypeptide chain by a single type of amino acid with probability close to 1 (conservation) can be viewed as a spontaneous breaking of amino acid content (second order- like phase transition\cite{Ram:94} (see also \cite{Bry:89}) corresponding to the transition between random chains and good folders\footnote{Second order phase transitions are, as a rule, connected with changes in symmetry. Consequently, the two phases cannot coexist (e.g. aligned (ferromagnetic) and non--aligned (paramagnetic) phases). Starting at a critical temperature, the new phase grows continuously. This is at variance with first order phase transitions (e.g. denatured $\rightarrow$ native, taking place in the case of the folding of proteins), where, in the thermodynamic limit, the two phases can coexist (discontinuous changes of the order parameter between the two phases). In particular, at the folding temperature $T_f$ the probabilities that the system is in the native and in the denatured states are equal.}). The most important emergent property\footnote{That is properties not present in the Hamiltonian describing the system, neither in the ``particles'' (amino acids) forming it. In the case of paramagnetic$\rightarrow$ferromagnetic phase transition emergent properties are, for example, domain walls, magnetic rigidity, etc\cite{And:72,And:95}.} associated with this, symmetry--breaking, second order--like phase transition is folding (first order phase transition; coexistence of native\footnote{Good folders (and consequently hot amino acids and LES) lead to native states which can be viewed as scale free networks\cite{Bar:02,Ven:02}.} (N) and denatured (D) states, see Fig. \ref{fig:protevo}(e)), a phenomenon tantamount to biological activity and eventually to metabolic function and thus to the emergence of life on earth.

\begin{figure*}[htb!]
\begin{center}
\includegraphics[width=0.82\textwidth]{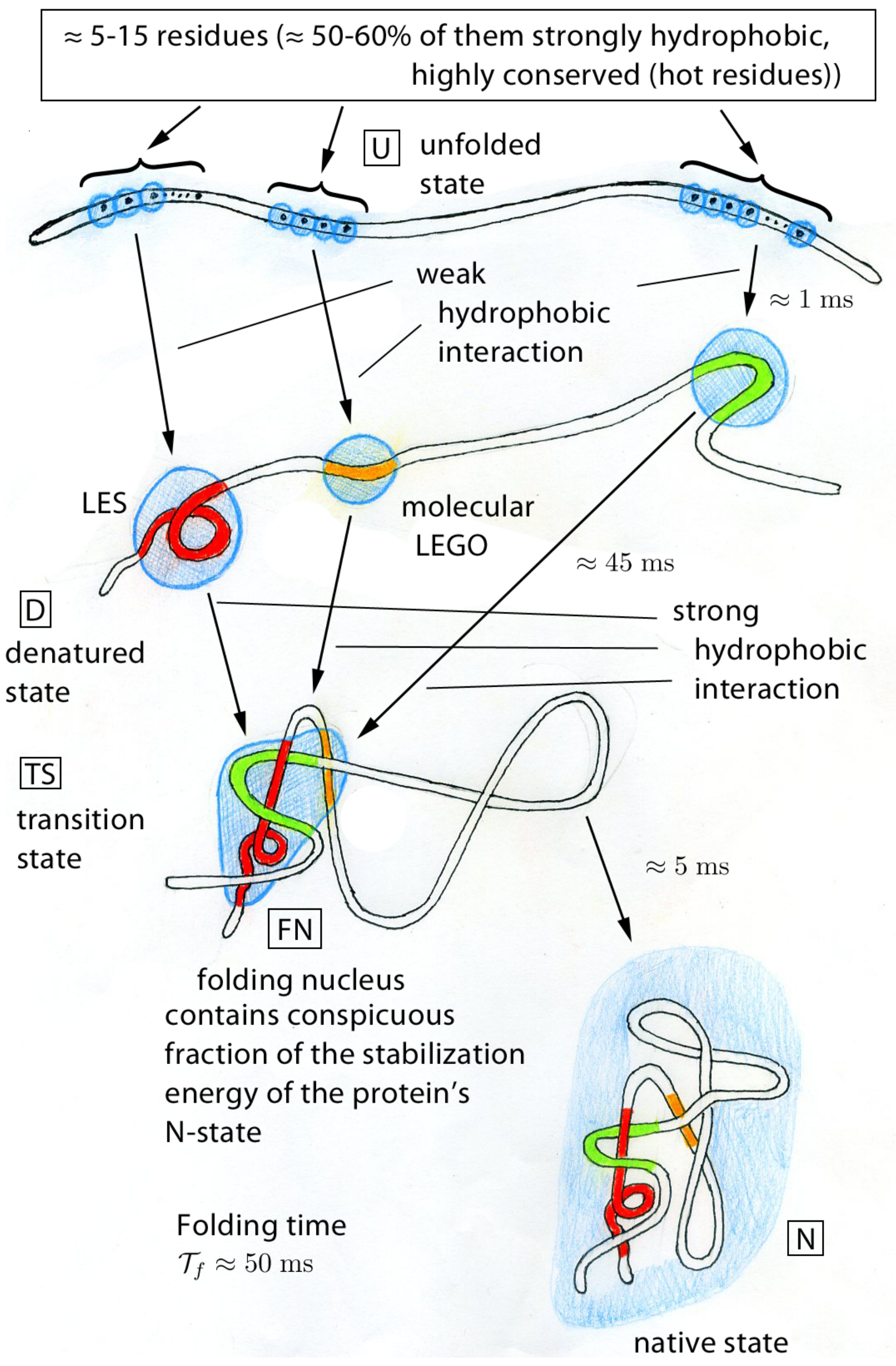}
\caption{ }
\label{fig:hydro}
\end{center}
\end{figure*}

\begin{figure*}[htb!]
\begin{center}
\includegraphics[width=0.87\textwidth]{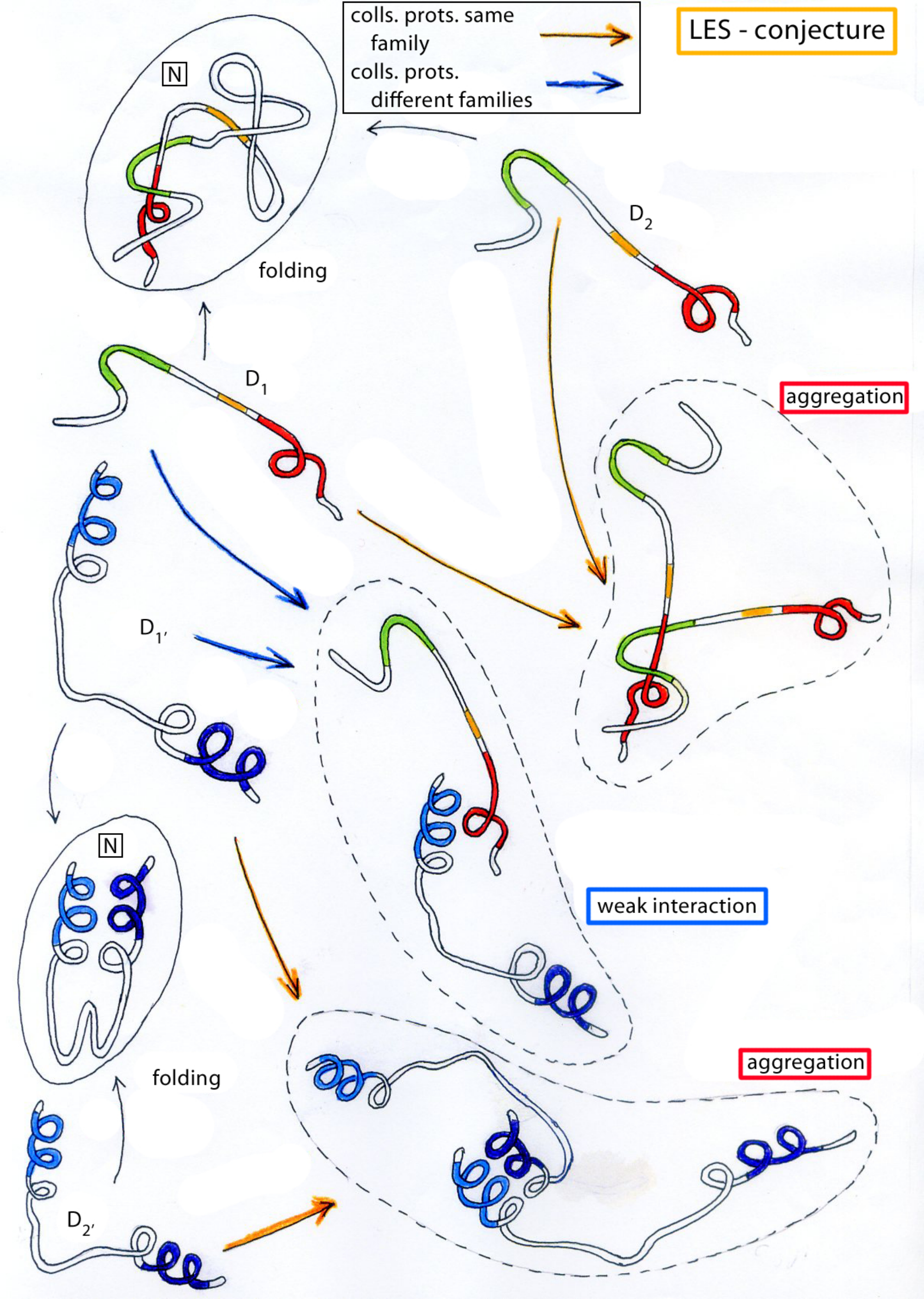}
\caption{ }
\label{fig:lesconj}
\end{center}
\end{figure*}

LES, which can be viewed as incipient, virtual secondary structures already present with varied stability in the denatured state, control not only folding (Fig. \ref{fig:hydro}), but also aggregation\cite{Bro:98,Sha:99}. Because folding is much faster than collision events between solvent exposed LES of different proteins belonging to each group of proteins of a given type present in the cell ($\approx 10^2$), but much slower than collisions between the solvent exposed LES associated with all of the $10^6$ proteins belonging to a cell\cite{Ful:82,Goo:91,ECo:www}, one is essentially forced to assume that evolution has tooled the LES of each protein to recognize their complementary (like left and right hands) LES and essentially nothing else (\textbf{LES--conjecture}) (see Fig. \ref{fig:lesconj}).
This conjecture is consistent, among other things, with the dearth of folds revealed by the proteomic project\cite{Ser:05}, project aimed at determining the native conformations of all human proteins.

\begin{figure*}[htb!]
\begin{center}
\includegraphics[width=0.97\textwidth]{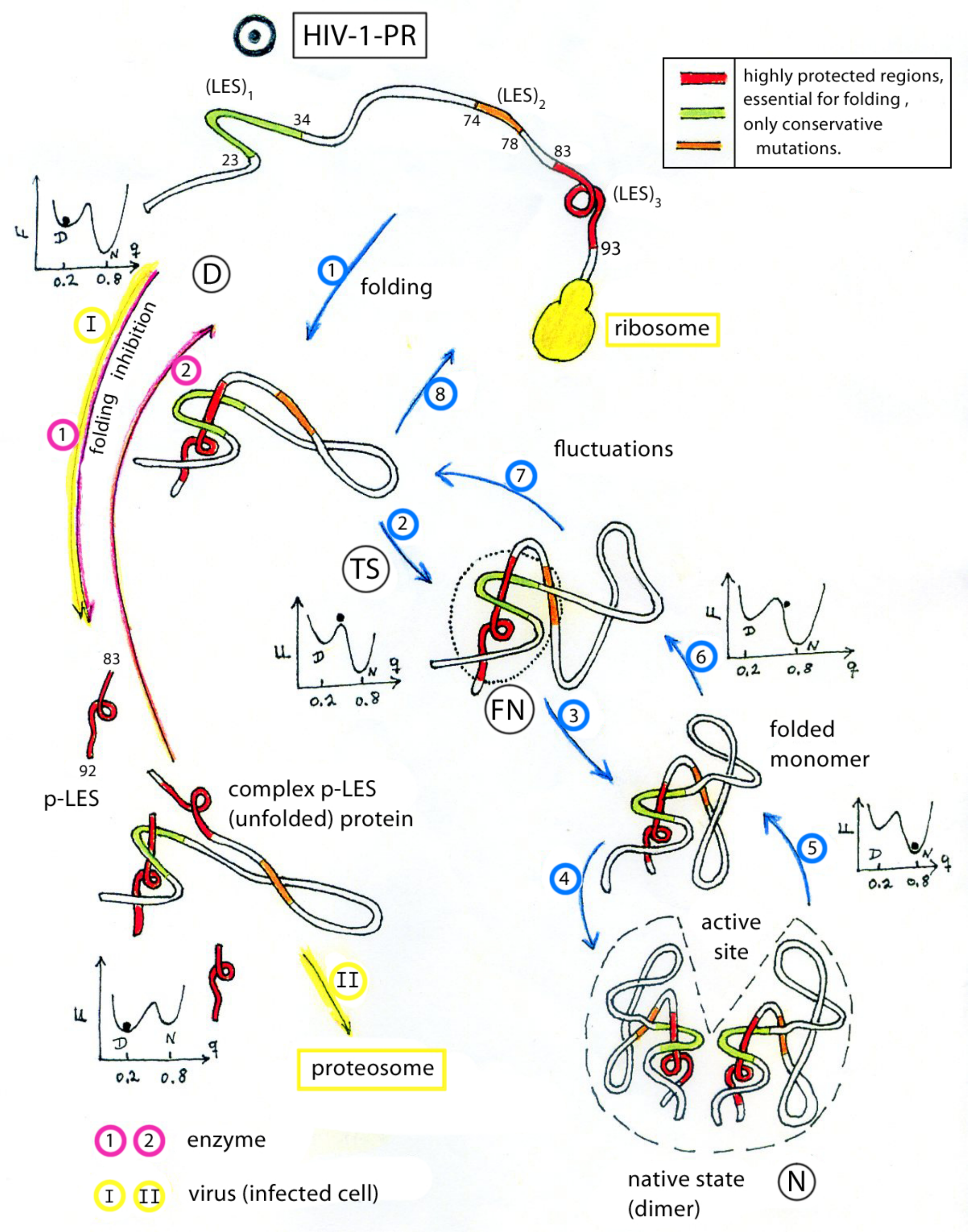}
\caption{ }
\label{fig:foldpep}
\end{center}
\end{figure*}

To make virtual LES become real, one can intervene the folding process with peptides displaying identical sequence of a LES of the protein under study\cite{Bro:03} (Fig. \ref{fig:foldpep} as well as App. D).
Such peptides, called p--LES, can bind a complementary LES leading to misfolding and thus competing with productive folding\cite{Bro:98,Pin:92,Bor:11}. Circular dichroism is consistent with such a scenario\cite{Bro:06b,Bro:05}, while NMR indicates that the only amino acids which give a signal close to that associated with the native state of the protein are those which bind in the native state to the LES of which the peptide p--LES is a replica\cite{Cal:09}.

\begin{figure*}[htb!]
\begin{center}
\includegraphics[width=0.93\textwidth]{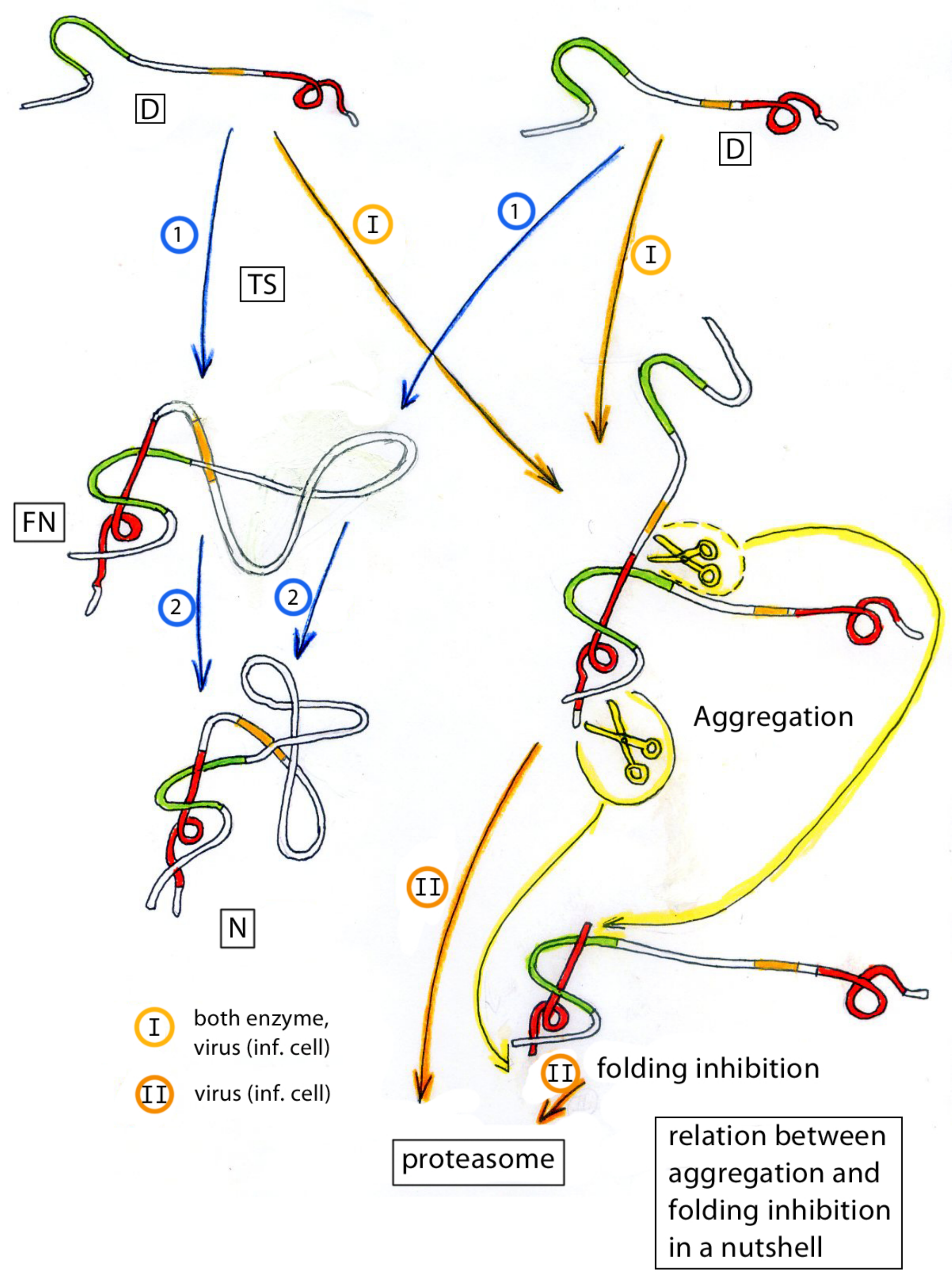}
\caption{ }
\label{fig:foldinh}
\end{center}
\end{figure*}

\begin{figure}[htb!]
\begin{center}
\includegraphics[width=0.49\textwidth]{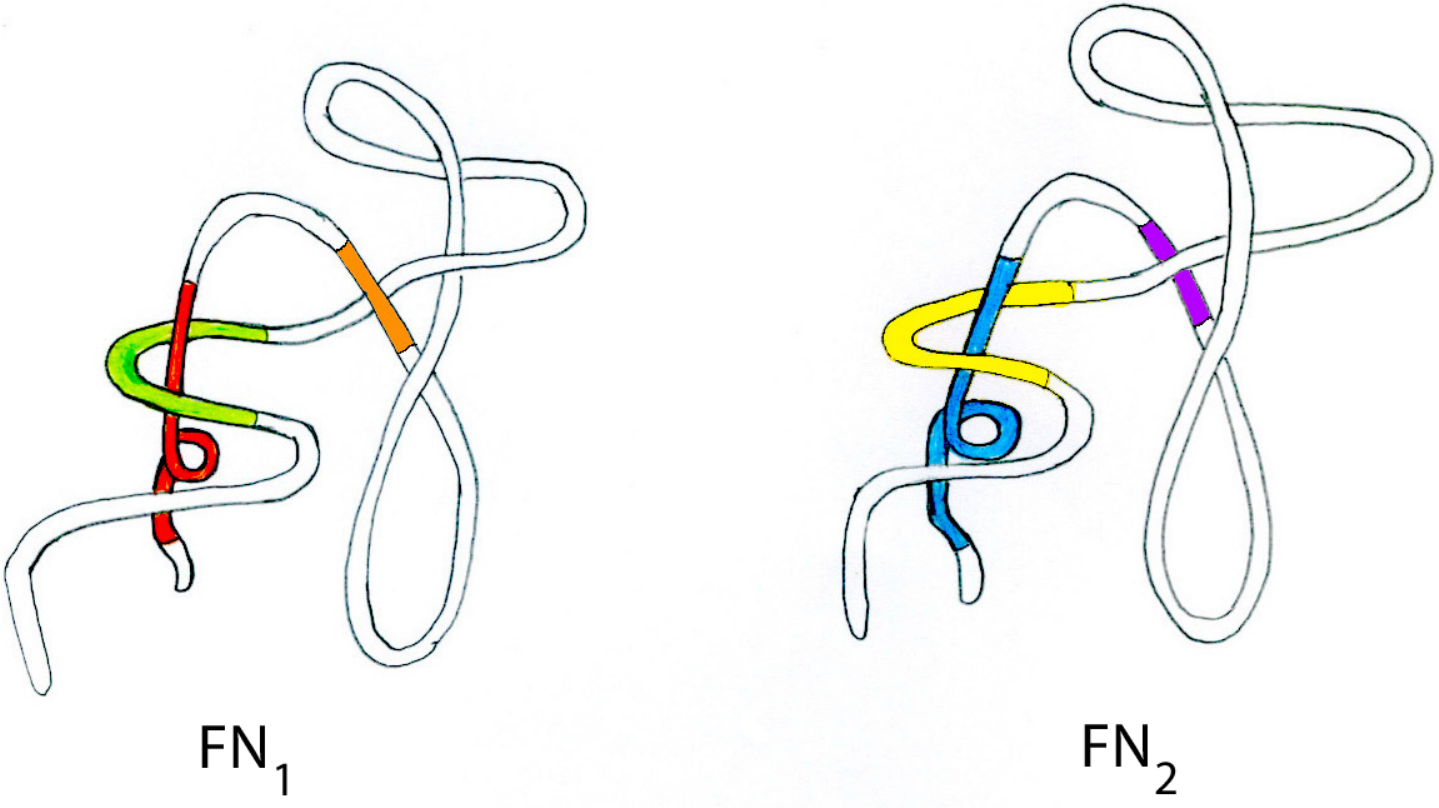}
\caption{ }
\label{fig:twofns}
\end{center}
\end{figure}

This insight concerning the validity of the LES--conjecture can be used at profit to understand aggregation, to design novel enzymes as well as help solving the protein folding problem: for this purpose one should design all possible LES$\rightarrow$FN$\rightarrow$folds and establish the connection between amino acid sequence and LES (3 steps strategy, cf. ref.\cite{Bro:01b}). Although this is a central issue in the study of proteins, arguably, the most promising role of p--LES is that of being leads to non--conventional (folding) inhibitors (Fig. \ref{fig:foldinh}), drugs likely not to create resistance. In fact, the only way a target protein can avoid that a p--LES binds to its complementary LES is by mutating the hot amino acids of its LES. But such an event will lead to denaturation. This does not mean that a target protein cannot develop resistance. It only means that to do so a concerted mutation of a large number of amino acids has to take place in a single step, an event which is very unlikely\cite{Tia:09}. In fact, there are no point--mutation--paths connecting the (few) possible FN of a protein (Fig. \ref{fig:twofns}; see also \cite{Tia:00}).

From this vantage point of view, an eventual confirmation of the validity of the LES--conjecture would imply the emergence of a new paradigm in the design of drugs and thus in the cure of diseases, in particular infectious diseases: (high mutation barrier) folding inhibition\footnote{``First we guess it. Then we compute the consequences of the guess to see what would be implied if the law we guess is right. Then we compare the results of the computation to nature, with experiment or experience, compare it directly with observation, to see if it works. If it disagrees with experiment it is wrong. In that simple statement is the key to science. It does not make any difference how beautiful your guess is. It does not make any difference how smart you are, who made the guess, or what your name is -- if it disagrees with experiment it is wrong. That all there is to it.'' \emph{R. P. Feynman}}. Furthermore, these drugs should display little side effects in keeping with the LES--conjecture (see also Fig. \ref{fig:lesconj}). In the case of the HIV--1--PR such advantage should also carry to the low toxicity for the proteasome.

To shed light on this issue, assuming the target protein to be an enzyme like e.g. the HIV--1--Protease, folding inhibition and thus loss of activity has to be measured in vitro (purified enzyme\cite{Bro:06b}), in acute and in chronically infected cells (virus\cite{Rus:07,LoC:07}), in vitro passage over long periods of time (\cite{Fer:08,Fer:09}) and in living organisms, that is in test animals first and in AIDS patients later\footnote{Tests in animals (pharmacokinetics, pharmacotoxicity) and in patients (clinical phase I and phase II and eventually III) are very expensive, running into the millions of euros the first, and in tens if not hundreds of millions the second. And this kind of money can only be provided by big pharm, provided one has deposited patent requests before publishing the results of basic research (\cite{Bro:pat,Bro:patUS}). There are many ways one can interact with pharmaceutical firms to have such very expensive experiments carried out. A very attractive one (in paticular if one has the luck to hit on a firm willing to fully support cutting edge research without strings attached) is through an University--pharmaceutical spin--off like Foldless S.r.l.\cite{Foldless:news}.}. The wonder of an eventual positive outcome of these tests is that of helping eradicate one by one some of the worst flagels afflicting humanity. At the end, this (\textbf{if it cures it is right}) would be the real outcome of the \textbf{LES}--conjecture, which infinitely trascends that of Feynman (if it agrees with the experiment it is, if not right at least not wrong) in spite of its brilliancy. All the work, let alone investments, poured in the last decade to map out the consequences and develop embodiments of the spontaneous breaking of amino acid symmetry content phenomenon, will be wholly justified by the cure of even a single infected person.


\vspace{1cm}
Figure \ref{fig:protevo}: Schematic representation of protein evolution starting from short ($\approx 30$ aa long) random polymers (a) to enzymes (homodimer) (e). It is of notice that a similar scenario is obtained by discussing protein evolution (folding domains) in terms of single, $\approx 100$ amino acid long chains. Likely, both paths were tried by nature (within this context see App. C, as well as \cite{Pan:96}). In drawing the cartoons one had in mind the HIV--1--protease, a dimer made out of two identical chains (homodimer) each containing 99 amino acids (this is also true for the other figures, note however the variance in connection with Fig. \ref{fig:hydro}).

\vspace{0.3cm}

Figure \ref{fig:hydro}: Schematic representation of the role played by (weak and strong) hydrophobicity\cite{Cha:05,Cha:02} in the folding of a protein (see also \cite{Ban:07} and App. B). It is of notice that the times shown can be considered typical for 100 residue long proteins \cite{Fer:99}, but not for the HIV--1--PR monomer, which folds in times of the order of seconds \cite{Noe:09}.

\vspace{0.3cm}

Figure \ref{fig:lesconj}: LES--conjecture. The cell is crowded with $\approx 10^6$ proteins ($\approx 10^2$ of each type) To be able to fold, LES of one type of proteins must recognize its complementary LES and nothing else (see App. D).

\vspace{0.3cm}

Figure \ref{fig:foldpep}: Schematic representation of the intervening of a process of folding with peptides (p--LES) displaying the same sequence as one of the LES of the target protein.

\vspace{0.3cm}

Figure \ref{fig:foldinh}: Schematic representation of the link existing between LES, folding and aggregation\cite{Bro:98}, which is at the basis of (non--conventional) folding inhibitors.

\vspace{0.3cm}

Figure \ref{fig:twofns}: Schematic representation of the two folding nuclei\cite{Tia:09,Cha:11} expected for the HIV--1--PR. Different colors mean different amino acid sequence.

\newpage


\clearpage
\newpage

\appendix



\section{Hierarchical folding}

The fact that the \textbf{L}ocal \textbf{E}lementary \textbf{S}tructure (\textbf{LES}) ---scenario (--conjecture) under different guises (foldons, folding initiation sites, nucleation centers, hydrophobic initiation sites, etc), has been discovered again and again, testifies to its soundness and universality. In what follows, paradigmatic examples of such, almost repetitive although in most cases independent ``eureka''--like events, spanning half a century of research (1962--2011), are briefly discussed, starting with work done in the sixties by Anfinsen and collaborators (see \cite{Hab:62,Eps:63,Sac:72,Anf:73,Mat:78,Mer:85,Pin:92,Cot:90,Onu:95,Wal:98,Pan:96,Nav:01} \cite{Cai:02,Han:98,Mai:05,Dys:06,Har:11,Sos:11,Lin:11,Sti:11}, and refs. therein).

\textbf{Biological function} appears to be more a correlate of \textbf{macromolecular geometry} than of chemical details\footnote{Protein folding, basic physics more than detailed chemistry}, and considerable modification of protein sequence may be made without loss of function. Mutations and natural selection are permitted with a \textbf{high degree of freedom} (\emph{cold sites})\cite{Tia:98} during accidental mutation, but a limited number of residues (\emph{hot and warm sites})\cite{Tia:98,Bro:98}, destined to become involved in the internal, hydrophobic core of proteins (\emph{FN}), must be \textbf{carefully conserved}\footnote{Note however ref. \cite{Lar:02}. It is of notice that, making use of the definition of $\phi$--value\cite{Fer:99}, namely $\phi = \left( \Delta G_W^{TS-D} - \Delta G_M^{TS-D} \right)/\left( \Delta G_W^{N-D} - \Delta G_M^{N-D} \right)$, and of the fact that the partial ``flickering'' in and out of the LES, stabilized at various degrees by the hot residues, one would get for these residues ratios displaying essentially any value, in keeping with the fact that both numerator and denominator are small.}, or at most replaced with other residues which display a close chemical and physical similarity, let alone hydrophobicity (conservative mutations). Only the geometry of the protein and its active site need be conserved, except for such residues as actually parti\-ci\-pate in a unique way in a catalytic or re\-gu\-latory mechanism.

Because a chain of 99 amino acid residues with two rotable bonds per residue, each bond having two or three permissible or favored orientations, would be able to assume on the order of $4^{99}$ to $9^{99}$ different conformations in solution, it is necessary to postulate the existence of a limited number of allowable initiating events (\textbf{nucleations}) in the folding process\cite{Lev:69}, essentially controlled by \textbf{hydrophobic forces} (see App. B). This is in keeping with the fact that in aqueous solution, ionic and hydrogen--bonded interactions would not be expected to compete effectively with interactions with solvent molecules and anything less than a \textbf{sizeable nucleus of interacting amino acid} side chains ((hC), (TS), (FN), (FN)$_\textrm{pc}$)\footnote{(hC): hydrophobic core, (TS): transition state, (FN): folding nucleus, (FN)$_\textrm{pc}$: post critical (FN)} would likely have a very short lifetime\footnote{``$\ldots$ Furthermore, it is important to stress that the amino acid sequences of polypeptide chains designed to be the fabric of protein molecules only make functional sense when they are in the three--dimensional arrangement that characterizes them in the native protein structure'' (p. 228, ref. \cite{Anf:73}, third column). In this statement one finds all of what eventually became known as inverse folding problem (and associated solution, see refs. \cite{Sha:94,Tia:98,Bro:98,Bro:01b} and refs. therein) and as the G\=o--model\cite{Go:83}. On the actuality of such a statement expressed in \textbf{1973}, one can read the comment ``How proteins fold'' (ref. \cite{Sos:11}, published in \textbf{2011}): ``Non native structure has minimal influence on the (folding, rab) pathway. If non--native contacts are also insignificant, the widespread use of G\=o--models $\ldots$ would be justified.'' (p. 465, third column). See also refs. \cite{Lin:11} and \cite{Sti:11}.}. It seems reasonable to suggest that portions of a protein chain that can serve as \textbf{nucleation sites} for folding will be those that ``flicker'' in and out of the conformation they occupy in the final protein (complementary LES) and which, upon docking, will form a relatively rigid structure\footnote{Using antibodies which can recognize structured segments of staphylococcal nuclease suggests that e.g. approximately 0.02 percent ($2\times 10^{-4}$) of fragment 99--149 exists in the native format at any moment. Such a value, although low, is probably very large relative to the likelihood of a peptide fragment of a protein being found in its native format on the basis of chance alone\cite{Anf:73}}, stabilized by a set of cooperative interactions. These \textbf{nucleation centers}, in what has been termed their ``\textbf{native format}'', might be expected to involve such potentially self--dependent substructures as helices, pleated sheets, or beta--bends\footnote{Because these elements (foldons\cite{Mai:05,Pan:96}, nucleating (hydrophobic) pockets\cite{Mat:78}, folding initiation sites\cite{Dys:06}, transient local structures\cite{Mat:78}, hydrophobic folding units, partially folded kinetic intermediates, specific subdomain structures\cite{Hab:62,Eps:63,Sac:72,Mat:78,Mer:85,Onu:95,Wal:98,Pan:96,Nav:01,Cai:02,Han:98,Mai:05,Dys:06,Har:11,Sos:11,Lin:11,Sti:11}, nucleation sites\cite{Anf:73,Pin:92,Mat:78}, etc) often are intrinsically unstable, low--energy pathways are likely to involve foldons building on top of existing structures in a process of sequential stabilization (ref. \cite{Sos:11}, p. 465, bottom second column)}.

The examples of noncovalent interaction of \textbf{complementing fragments}\cite{Anf:73,Mer:85} referred to in connection with protein evolution gives strong support to the idea that, at the basis of protein folding, there appears to exist a very fine balance between stable, native protein structure (even with low but much larger probability than that of random sequences) and random, biologically meaningless polypeptide chains.

As an example let us refer to the fragment 1--126 of staphylococcal nuclease molecule cited in ref. \cite{Anf:73}. This fragment contains all of the residues that make up the active center of nuclease. Nevertheless, even if it represents about 85\% ($\approx$ 126/149) of the total sequence of the nuclease, it exhibits only about 0.12 percent of the activity of the native enzyme. The further addition of 23 residues during biosynthesis, or the addition in vitro, of residues 99--149 as a complementing fragment, restores the stability required for activity to this unfinished gene translation.

The process of folding can be shown to take place in at least two phases\footnote{The first phase is essentially temperature--independent (and therefore possibly entropically driven) the second being temperature--dependent (see ref. \cite{Anf:73}, p. 228 first column  and beginning second one). In keeping with Kramers, the transition rate from one phase to another is given by the relation $K \sim \exp^{- \frac{\Delta E}{T}}$. In the case in which there is no barrier between the two phases $\Delta E \rightarrow \Delta E_{eff} = -TS$ and $K \sim \exp^{\frac{TS}{T}} = \exp^{S}$. That is, the rate $K$ is temperature independent and thus entropically driven. In other words, there are no barriers to be overcome to find the new phase. The polypeptide chain undergoes a random walk in conformation space until it clicks on the conformation characteristic of such a phase. Assuming contact interactions the system will acquire the needed stability to eventually undergo the second, enthalpic driven and thus temperature dependent phase, in which the system has to overcome a potential barrier. If, however, the force driving the system is finite range, like e.g. the hydrophobic force, then the clicking and stabilization on the conformation characteristic of the first phase has to take place under the entropically driven WHI component of this interaction (for groups of hydrophobic residues which can fit together in a compact conformation of radius $\lesssim 1$ nm, it is the volume dependent, entropic controlled \textbf{W}eak \textbf{H}ydrophobic \textbf{I}nteraction which stabilizes the system (or the different groups),see App. B, in particular Fig. B.1 and B.2.}. An initial rapid folding with a half--time of about 50ms, and a second, somewhat slower transformation with a half--time of about 200ms. The first phase is essentially temperature--independent (on therefore possibly entropically driven, \textbf{WHI}) and a second temperature--dependent (\textbf{SHI})\cite{Cha:05,Cha:02}.



\section{Weak and strong hydrophobic interactions}

{
\setlength{\footnotesep}{0.8cm}

From Fig. B.1, corresponding to Fig. 2 (p. 642) of ref. \cite{Cha:05}, one can extract information concerning the Strong Hydrophobic Interaction (SHI) (see also ref. \cite{Cha:02}) proportional to the surface ($S = 4 \pi R^2$) of the hydrophobic cavity (solute)\footnote{Dissolving a substance in a solvent can be regarded as transforming a system from state \textbf{1 (pure solvent)} to state \textbf{2 (solvent plus solute)}. This process is associated with a change in the free energy $\Delta G = G_2 - G_1$, a quantity which is positive for hydrophobic solutes.}
$$ \Delta G_S^{\textrm{SHI}} \approx \gamma S = B S \;,$$
where
$$ B = \gamma \approx 7 \times 10^{-2} \textrm{ J}/\textrm{m}^2 \;,$$
is the liquid--vapour surface tension\footnote{The surface tension is a measure of the force that must be applied to surface molecules so that they experience the same force as molecules in the interior of the liquid. Surface tension exists because of attractive forces between the molecules in the bulk liquid (4 hydrogen--bonds in average) and the molecules in the surface ($\approx 2$ hydrogen--bonds in average).
\begin{wrapfigure}[10]{l}{0.27\textwidth}
\begin{center}
\vspace{-0.7cm}\includegraphics[width=0.22\textwidth]{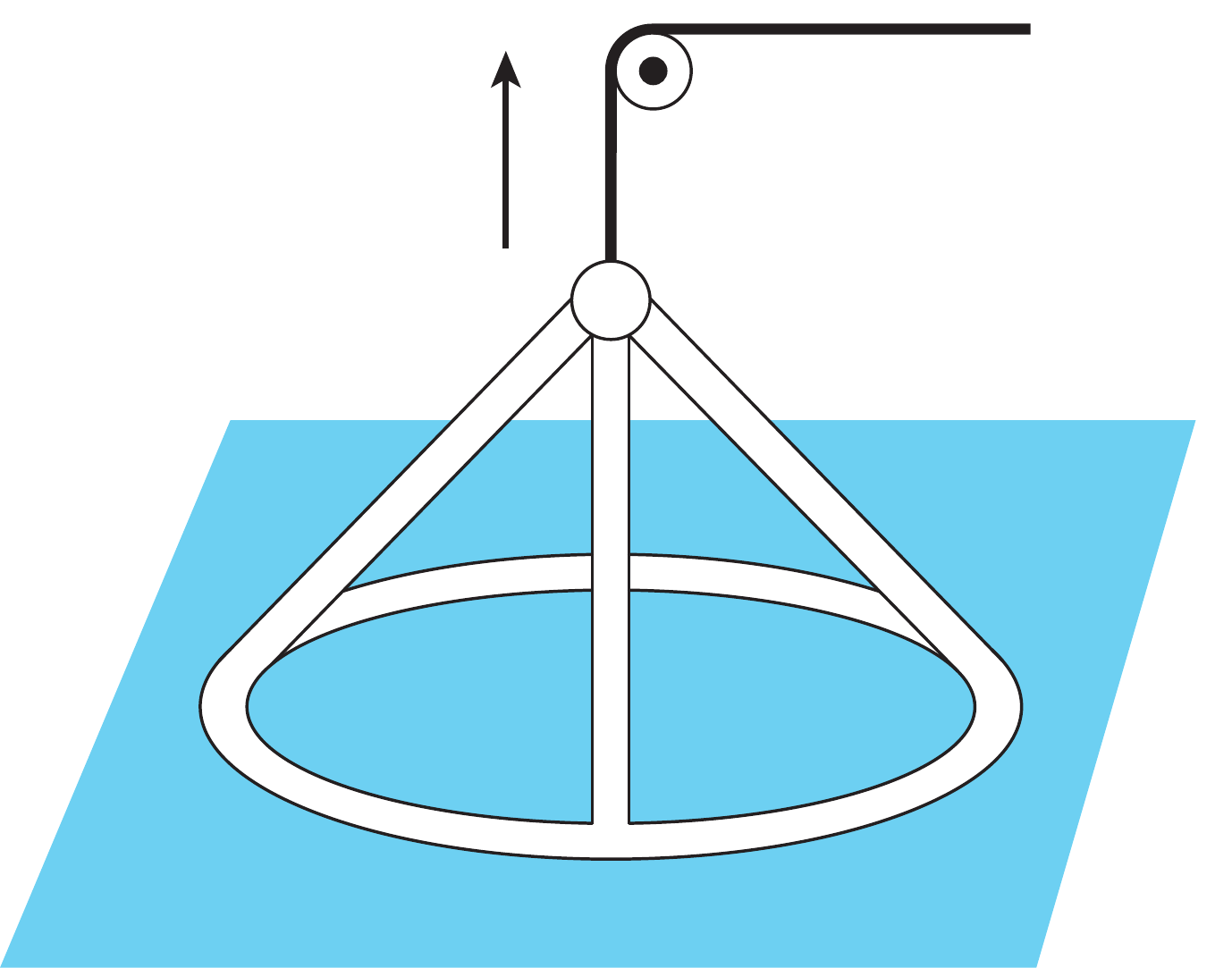}
\end{center}
\end{wrapfigure}
A molecule at the surface experiences a net inward force. This is the reason why a mosquito can walk on the water surface. Surface tension can be measured in standard experiments, by destroying part of the surface and recording how much work it takes to reconstruct it (see cartoon). At ambient conditions (room temperature and 1 atm pressure) liquid water lies close to phase coexistence with its vapour (in the figure it is shown a typical device used to learn about interaction controlling leptodermic systems is schematically shown).}.

\begin{figure}[hb!]
\begin{center}
\includegraphics*[width=0.47\textwidth]{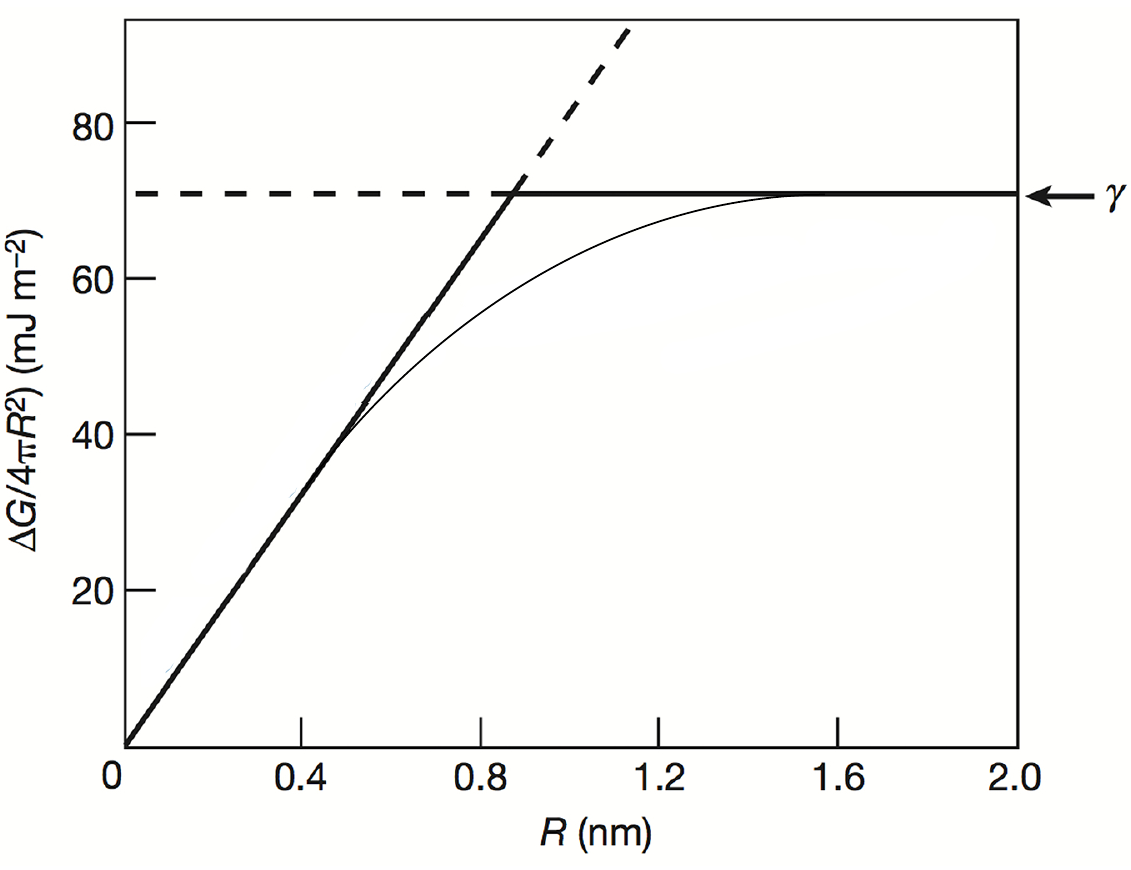}

{\footnotesize {Figure B.1: Schematic representation of the numerical results reported in ref. \cite{Cha:05} concerning the solvation free energy $\Delta G = G_2 - G_1$ for a spherical cavity in water normalized with respect to the surface area of the circumference resulting from the intersect of the cavity with a plane containing its center as a function of the cavity radius (room temperature and 1 atm of pressure). The liquid--vapour surface tension is denoted by $\gamma$ (\emph{$\Delta G > 0$ means that $G_2$ is less negative than $G_1$. In other words, immersing a non--polar molecule in water the free energy increases}).}}
\end{center}
\vspace{-0.4cm}
\end{figure}

\begin{figure*}[hb!]
\begin{center}
\includegraphics*[width=0.9\textwidth]{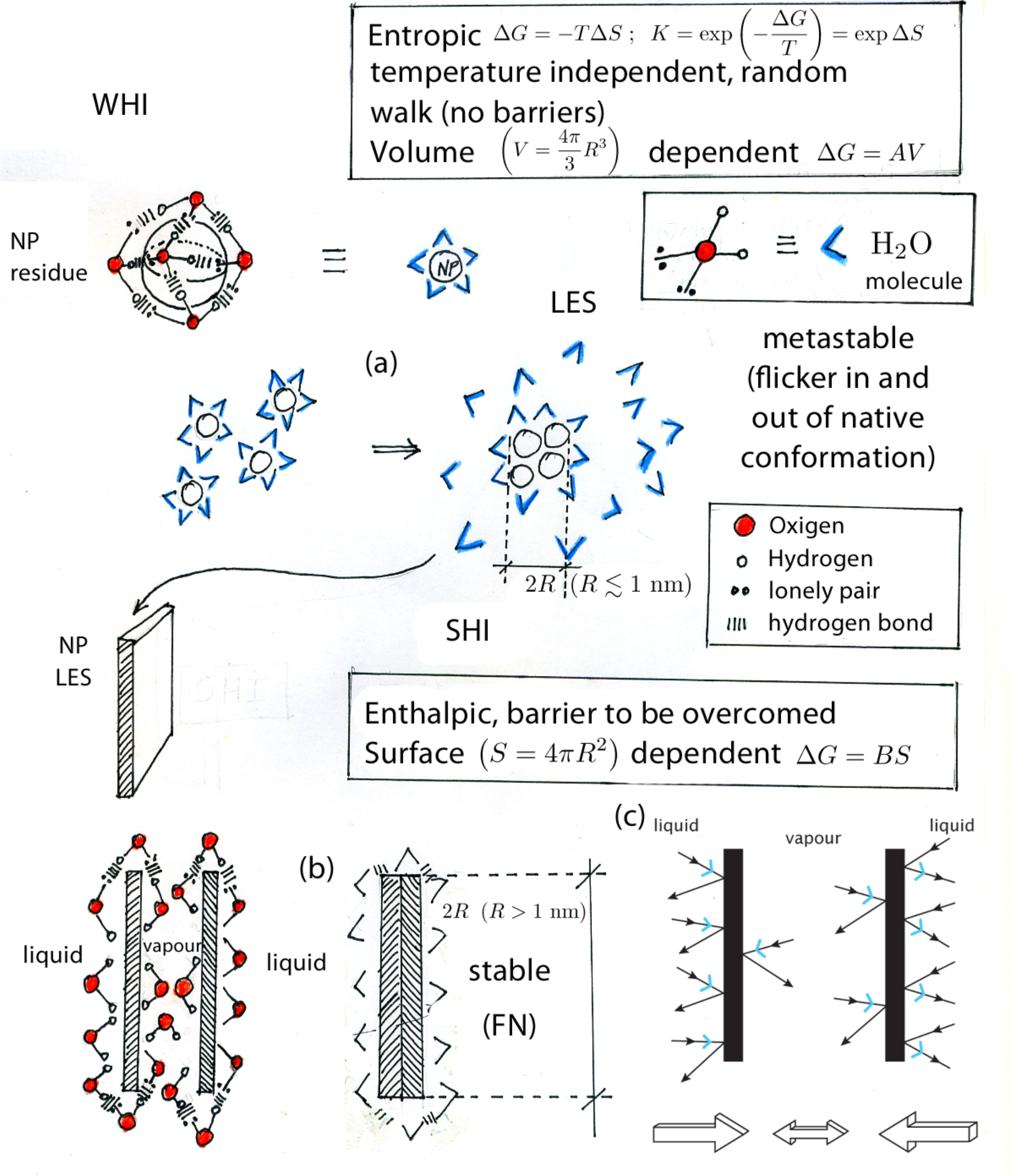}

{\footnotesize {Figure B.2: \textbf{(a)} Entropic, Weak Hydrophobic Interaction (WHI). Non--polar (NP) molecules of small dimension carry around solvation shells which preserve the hydrogen bonds of bulk water, although ``freezing'' the solvent molecules forced to go around the NP molecule. Assembling them together frees a number of water molecules, thus increasing entropy ($\Delta S > 0$). This mechanism, so called weak hydrophobic interaction, is operative for values of the radius of the conglomerate $\lesssim 1$nm. \linebreak \textbf{(b)} Enthalpic, Strong Hydrophobic Interaction (SHI). Large non--polar molecules or clusters of NP molecules ($R>1$nm) represented by extended plates, impedes solvent molecules to form hydrogen bonds as in bulk water. The number of these molecules is drastically reduced, with a net gain of enthalpy, by overlapping the two plates. \textbf{(c)} Of notice that at biological conditions (room temperature and 1 atmosphere pressure), water is essentially in equilibrium with its vapour, the phase essentially found between plates. Fluctuations are able to expel these molecules, thus reducing the internal pressure and allowing the plates to come into contact under the pressure of the, external, water molecules. One can thus view the effect of SHI on NP molecules as a generalized Casimir effect \cite{Cas:48,Cas:48b}.}}
\end{center}
\vspace{-0.4cm}
\end{figure*}

From Fig. B.1 one observes that the calculated value of $\Delta G/4 \pi R^2$ is equal to 50 mJ/m$^2$ for $R=0.6 \textrm{ nm} \;(1 \textrm{ nm} = 10^{-9} \textrm{ m} = 10 \textrm{ \AA})$, in the region in which $\Delta G$ coincides with $\Delta G_V^{\textrm{WHI}}$. Thus
$$ \frac{\Delta G_V^{\textrm{WHI}}}{4 \pi R^2} = \frac{\Delta G_V^{\textrm{WHI}}}{V} \frac{R}{3} = 50 \times 10^{-3} \textrm{ J}/\textrm{m}^2 \;, \;\;(R=0.6 \textrm{ nm}) \;, $$
$$ \frac{\Delta G_V^{\textrm{WHI}}}{V} = \left( \frac{3 \times 50}{0.6 \textrm{ nm}} \times 10^{-3} \textrm{ J}/\textrm{m}^2 \right) \;.$$
Then
$$ \Delta G_V^{\textrm{WHI}} = AV \;,$$
where
$$ A \approx 3 \times 10^8 \textrm{ J}/\textrm{m}^3 \;. $$
The crossing radius, that is the value of the radius of the hydrophobic particle or group of particles (see Fig. 1 ref. \cite{Cha:05}) which marks the transition from the volume dependent, entropic WHI regime, to the surface controlled, enthalpic SHI situation (see also Fig. B.2) is determined by the relation
$$ \Delta G_V^{\textrm{WHI}}(R_{\textrm{cross}}) = \Delta G_S^{\textrm{SHI}}(R_{\textrm{cross}}) \;, $$

\noindent that is
$$ AV = BS \;. $$
Making use of the relation
$$ \frac{V}{S} = \frac{\frac{4\pi}{3} R^3}{4 \pi R^2} = \frac{R}{3} \;,$$
and
$$ \frac{B}{A} = \frac{7 \times 10^{-2} \textrm{ J}/\textrm{m}^2}{3 \times 10^8 \textrm{ J}/\textrm{m}} = \frac{7}{3} \times 10^{-10} \textrm{ m} \;,$$
one obtains
$$ \frac{R_{\textrm{cross}}}{3} = \frac{7}{3} \times 10^{-10} \textrm{ m} \;, $$
that is,
$$ R_{\textrm{cross}} \approx 0.7 \textrm{ nm} \;. $$
In other words, the change of regime corresponds to a radius of the order of 1 nm (=10 \AA). Making use of the fact that the average Van der Waals volume of the 20 most common amino acids is $\approx 120$ \AA$^3$ ($\Rightarrow R_{aa}(\textrm{VW}) \equiv R_{aa} \approx 3.1$ \AA) and that the average range of the associated (attractive) interaction is $\approx 0.5$ \AA -- 1 \AA, one expects that the average Wigner cell radius of an amino acid is $R_{aa} \approx 4.5-5$ \AA. From this estimate one obtains that the number $n \; \left( \approx \left( R_{\textrm{cross}}/R_{aa} \right)^3 \right)$ of amino acids which can fit into the largest hydrophobic cavity which immersed in water does not deplete hydrogen bonds (WHI) is bound between the values $\left( 10 \textrm{ \AA}/ 4.5 \textrm{ \AA} \right)^3$ and $\left( 10 \textrm{ \AA}/ 5 \textrm{ \AA} \right)^3$, that is
$$ 8 \lesssim n \lesssim 11 \;. $$
It is strongly suggestive that this number is similar to that corresponding to the number of amino acids forming LES of typical globular proteins. This is even more so if one takes into account the fact that hydrophobicity in connection with amino acids refers principally to side chain transfer experiments (vapour $\rightarrow$ liquid water) ignoring the hydrophobic parts of the backbone.

Of notice that the hydrophobic (WHI) force is likely to lead to metastable (see also \cite{Hua:01}), strongly fluctuating, LES (structuring probabilities $\approx$ few \%; within this context see ref. \cite{Anf:73}) in keeping with their size $<1$ nm, and only to sizable stabilities (SHI) of the (postcritical) FN.


\setcounter{section}{2}

\section{Enzymatic retention of native structural ``memory''. Protein evolution in terms of few ``short'' segments}

The \textbf{124--residue} bovine pancreatic ribonuclease A \linebreak (RNase A) has been extensively studied, let alone chemically synthesized. It was found, among other things, that the C--terminal peptide 111--124, the N--terminal peptide 1--20, and the central protein component 21--118 could be mixed together non--covalently and ribonuclease activity would be generated\cite{Anf:73,Pin:92,Mer:85}.

Other examples of retention of native structural ``memory'' by segments of a protein, have been found with complexing fragments of the staphylococcal nuclease molecule, a calcium--dependent, RNA-- and DNA--cleaving enzyme containing \textbf{149 amino acids} and devoid of disulfide \linebreak bridges and sulphydryl groups. The protein is digested by proteolytic enzymes. Tripsin, for example, cleaves the staphylococcal nuclease enzyme at a number of sites. \emph{The resulting fragments (residues 6 to 48) and (49 to 149) or (50 to 149) are devoid of detectable structure in solution}. However, as in the case of ribonuclease S, when the fragments are mixed in \emph{stoichiometric} amounts, regeneration of activity, about 10\%, and of native structure characteristics occurs\cite{Anf:73}, the complex being known as nuclease T.


\section{Folding inhibition}

As stated in ref. \cite{Anf:73}, methods that depend on hydrodynamic or spectral measurements are not able to detect the presence of the flickering in and out of the native conformation of the nucleation sites (virtual processes). On the other hand, a method which can reveal such events and which was employed in a study of the folding of staphylococcal nuclease and its fragments, is based on specific \textbf{antibodies against restricted portions of the amino acid sequence}\footnote{Of notice that this idea translates the sequence based strategy developed in ref. \cite{Bro:07} (see also \cite{Bro:pat}) from peptides to antibodies (Eugenio Cesana, private communication). Within this context cf. \cite{Bro:unp}.}.

\clearpage

Antibodies against specific regions of the nuclease \linebreak molecule were prepared by immunization of goats with either polypeptide fragments of the enzyme or by infection of the intact, native protein with subsequent fractionation of the resulting antibody in correspondence with the protein segments of interest.

In the former manner an antibody was prepared specifically directed against the polypeptide, residues 99--149, known to exist in solution as a random chain without the extensive helicity that characterizes this portion of the nuclease chain when present as part of the intact enzyme. Such an antibody preparation is referred to as anti--(99 to 149){\boldmath $_{r}$}, the subscript indicating the, random, disordered (denatured) state of the antigen.

A similarly specific antibody for the sequence (99 to 149) was obtained but this time by fractionation of antiserum to native nuclease. While this fraction, termed anti--(99 to 149){\boldmath $_{n}$} exhibited a strong inhibitory effect on the enzymatic activity of nuclease, anti--(99 to 149){\boldmath $_{r}$} was devoid of such an effect. In keeping with this result, the subindex {\boldmath $n$} refers to the native format of this bit of sequence.

Similar inhibitory effects, but this time making direct use of polypeptides displaying identical sequence to segments of the target protein, were found in the case of the \textbf{124--residue} bovine pancreatic ribonuclease A (RNase A). The peptide His 105 -- Val 124, which forms in the native conformation of the protein a $\beta$--pleated sheet, completely inhibits the refolding of this protein at a concentration of 10 $\mu$M, from the reduced, denatured state at a 1:1 molar ratio of peptide to refolding protein. It has also been observed complete inhibition of refolding by peptides 11--31 and 40--61, but this time at concentrations 100 $\mu$M and $>100$ $\mu$M respectively.

The basis for a possible explanation of these observations is the fact that, if a segment of a protein adopts a native--like conformation as an isolated peptide, it may inhibit protein refolding if this segment of the protein is involved in folding in an early stage of the refolding process. Inhibition would result from competition of the exogeneous peptide with its counterpart in the protein for interacting with complementary regions of the refolding protein (cf. ref. \cite{Pin:92}, see also \cite{Bro:08}).


\bibliographystyle{epj}


\nocite{Sos:11}
\nocite{Lin:11}
\nocite{Sti:11}
\nocite{Bro:07}




\end{document}